\documentstyle[11pt]{article}

\def\beqn{\begin{eqnarray}}
\def\eeqn{\end{eqnarray}}
\begin{document}

\title{{\bf Chiral Lagrangians and the QCD String}}
\author{
{{\sc J.Alfaro}\thanks{%
jalfaro@chopin.fis.puc.cl}} \\
{Departamento de F\'\i sica}\\
{Universidad Cat\'olica de Chile, Santiago, Chile}\\
\\
{{\sc A.Dobado}\thanks{%
dobado@eucmax.sim.ucm.es}} \\
{Departamento de F\'\i sica Te\'orica} \\
{Universidad Complutense de Madrid, Madrid, Spain}\\
\\
{{\sc D.Espriu}\thanks{%
espriu@ecm.ub.es}} \\
{Departament d'Estructura i Constituents de la Mat\`eria and IFAE} \\
{Universitat de Barcelona, Barcelona, Spain}\\
}
\date{}
\maketitle

\begin{abstract}
We propose a  method to derive the low-energy efective
action of QCD assuming that the long-distance properties
of strong interactions can be described by a string theory.
We bypass the usual problems related to the existence of the
tachyon and absence of the adequate Adler zero by using
a sigma model approach where excitations above the correct
(chirally non-invariant) QCD vacuum are included. Two-dimensional conformal
invariance then implies the vanishing of the ${\cal O}(p^4)$ effective
lagrangian coefficients. We interpret this result and discuss ways to go
beyond this limit.
\end{abstract}

\vfill

\clearpage

\section{Introduction}

There are many theoretical and empirical reasons that
make us believe that it should be possible to describe
QCD in terms of a string theory\cite{string}, at least
in some kinematical regime. The more commonly cited
arguments are the dominance of planar diagrams
in the large $N$ limit\cite{largeN} `filling
in' a surface (interpreted as the world-sheet of a string),
the expansion in terms of surfaces built out of plaquettes
in strong-coupling lattice QCD\cite{lattice}, and the success of
Regge phenomenology\cite{regge}, which can ultimately be understood
in terms of string theory ideas (although, as we will discuss a little
bit later, the actual Regge theory that corresponds to QCD cannot
be derived, at present, from any known string theory).

To these we could add two more reasons. One is the appeareance
in string theory of the universal (at least at long
distances) L\"uscher term\cite{ML}. The static interquark
potential provided by the string $V(r)=\sigma r + c$ gets modified
by quantum fluctuations by a Coulomb-like piece
$-\pi/12 r$, a term which come very handy when
fitting the string interquark potential to heavy quark spectra.
Finally, and in a completely different context, namely that of
deep inelastic scattering, the evolution of the parton distribution
down to low values of $Q^2$ (around (2 GeV)$^2$)
leads\cite{evol} to a low $x$ behaviour for the structure functions of the
form $x^{-1.17}$,
while Regge theory predicts $x^{-1}$, in striking good agreement.

Thus, that
there is a string description of QCD is
almost evident. Which is the appropriate string theory
for QCD, however? To answer this question one should
first ask oneself which is the kinematical regime
where the string picture would be applicable. It is
manifestly hard to reconcile the string picture and
high energy processes, such as deep inelastic scattering,
where the point-like structure
of quarks and gluons is apparent. While it is
quite conceivable that non-abelian gauge theories
could one day be understood in terms of a given string theory
(cf. the interesting recent developments about the
AdS/CFT relationship\cite{Mald}), it is
also quite obvious that this will never be the
natural language to understand high-energy processes.
We should probably be less ambitious and satisfy ourselves with
an effective description.

Not surprisingly, a lot of candidates
have been put forward as possible candidates of the
QCD string, ranging from the original Nambu-Goto
string\cite{NG} to the supersymmetric string\cite{LS}
and from the rigid string\cite{rigid} to the
five-dimensional string with manifest zig-zag
symmetry\cite{ZZ}. Most, if not all, of the
candidates are believed to be understood
as effective theories, valid only up to some
characteristic momentum transfer $k_{\rm max}$, and
at the (string) tree level, lest the inconsistencies
of string theory away from the critical dimensions
show up. We subscribe this point of view and think of strings as
effective theories and not worry at all about their
mathematical consistency as fundamental objects.

It is surprising that even with this modest and limited scope
all known string theories are inconsistent and cannot
provide a description of low energy QCD, even for
$k\ll k_{\rm max}$. To see how this comes about let us remember
the original Veneziano amplitude\cite{Ven}. After decorating
it with the appropriate Chan-Paton\cite{CP} factors\footnote{Due to
difficulties with unitarity only orthogonal groups can be introduced in this
way, but since there are other inconsistencies one should not worry too much
at this point.} it is supposed to describe the scattering amplitude of four
pions \beqn
{\cal A}(\pi^a \pi^b\to \pi^c \pi^d) \sim {\rm Tr}(T^a T^b T^c T^d) A(s,t)
+ {\rm non~cyclic~permutations},
\eeqn
\beqn
A(s,t)=
\frac{\Gamma(-\alpha(s))\Gamma(-\alpha(t))}{\Gamma(-\alpha(s)-\alpha(t))},
\eeqn
where $\alpha(s)=\alpha(0)+\alpha^\prime s$ is the Regge trajectory.
The inverse string tension $\alpha^\prime$ is to be determined latter.
In the Nambu-Goto string, from which this amplitude is deduced,
$\alpha(0)=1$.

In the above expression we immediately recognize
that there are poles in the $s$-channel whenever
$\alpha^\prime s= n-\alpha(0)$. Thus a tachyon is
present for $n=0$.

The supersymmetric string\cite{LS} does not really fare any better.
There are two sectors in supersymmetric strings. In the Neveu-Schwarz
(bosonic) sector $\alpha(0)=\frac{1}{2}$ and
there exist a (scalar/pseudoscalar) tachyon too. In Regge parlance
the spectrum in this sector is described by the `pion' trajectory
$\alpha_\pi(s)=\alpha(0) +\alpha^\prime s$, corresponding to
negative $G$-parity, and by the `rho' trajectory, corresponding
to positive $G$-parity,
$\alpha_\rho(s)=\alpha_\pi(s)+\frac{1}{2}$.
Usually one performs the
GSO projection\cite{scherk}, projecting out the tachyon. However one may
choose not to do so and compute the four-tachyon amplitude,
supposed to describe
pion-pion scattering, which is mediated by the exchange of particles in the
$\rho$-trajectory. The corresponding amplitude
is
\beqn
A(s,t)= \frac{\Gamma(1-\alpha_\rho(s))\Gamma(1-\alpha_\rho(t))}{
\Gamma(1-\alpha_\rho(s)-\alpha_\rho(t))},
\eeqn
or
\beqn
A(s,t)=
\frac{\Gamma(-s\alpha^\prime)
\Gamma(-t\alpha^\prime)}{\Gamma(-s\alpha^\prime-t\alpha^\prime-1)}.
\eeqn
This is the Lovelace-Shapiro amplitude, which contains no tachyonic poles.
Could it be a candidate to describe
pion scattering? The answer is no. It does not have the appropriate Adler
zero, i.e. the property that at $s=t=0$ the
amplitude vanishes.

A fix\cite{lovelace} to this problem is to replace {\em by hand}
$\alpha_\rho(s)$ by $\alpha_\pi(s)$. The amplitude becomes
\beqn
A(s,t)=
\frac{\Gamma(\frac{1}{2}-s\alpha^\prime)
\Gamma(\frac{1}{2}-t\alpha^\prime)}{\Gamma(-s\alpha^\prime-t\alpha^\prime)}.
\eeqn
with poles in the $s$-channel when $\alpha^\prime s= n +\frac{1}{2}$
It has no tachyons and the first pole is identified with the $\rho$
particle, thus fixing $\alpha^\prime$.
Furthermore, the previous amplitude has the right Adler zero. Based on
this amplitude
Polyakov and Vereshagin\cite{PV} have derived the first coefficients
of the effective chiral lagrangian and have found that
\beqn
L_1=\frac{1}{2} L_2,\qquad L_2=\frac{F_\pi^2}{8m_\rho^2}\ln 2, \qquad
L_3=-2L_2.
\eeqn
Numerically they turn out to be quite acceptable values\footnote{The
first relation, $L_1=L_2/2$, is a consequence of the large
$N$ (planar) limit.}, but unfortunately there is no way to justify the
apparently arbitrary change in the intercept.

An attempt to solve the difficulties associated to 
the presence of the tachyon is the formulation of the rigid
string\cite{rigid}, where four derivative interactions
contained in the second fundamental form modify
the string behaviour at short distances. Classically
at least, the Regge trajectories are modified\cite{Curtright},
making it conceivable that the tachyon is avoided. Unfortunately,
the classical trajectories are no more straight lines, something with ample
phenomenological support and thus a highly desirable property to preserve.
While the spectrum of the rigid string has not been determined at the quantum
level (the theory is not exactly solvable), it seems unlikely
at this point that it provides a satisfactory solution by itself,
even though it may be part of the solution, as we will later see.

It has been thought for a long time that the ultimate reason for the
presence of a tachyon in the spectrum lies in a wrong choice
of the vacuum\cite{tachyon}. Since the choice of the vertex operator,
$V(k)=:\exp{ikx}:$, is based on the Lorentz properties alone, it is the same
both for scalar and pseudoscalars and, accordingly, both scalars and
pseudoscalars have tachyonic poles in the $s$-channel on account of
parity conservation. The situation is thus parallel
to the one in multicomponent $\lambda \phi^4$ when
perturbing around $\phi=0$ gives negative $m^2$ values
for all components. It is natural
to assume that the amplitudes obtained through the use
of the canonical vertex operators correspond to
(unphysical)
amplitudes for excitations perturbed around the
wrong vacuum.

These ideas are certainly not new, but how could one
obtain the amplitudes for excitations around the
physically correct vacuum? We propose to use two
ingredients to try and give a partial answer
to this question.

The first one is to identify from the outset
the proper physical degrees of freedom. In this case, the
relevant degrees of freedom are the
ones emerging after the spontaneous breaking of
chiral symmetry. In the physical vacuum of QCD there
is a clear distinction between scalars (sigma particle) and pseudoscalars
(pions). The pseudoscalars can be collected in a
unitary matrix $U(x)$ which under chiral tranformations
belonging to $SU(3)_L\times SU(3)_R$
transforms as
\beqn
U(x)\to U^\prime(x) = L U(x) R^\dagger
\eeqn
$U(x)$ is nothing but a
bunch of couplings involving the variables $x$, from
the string point of view.

Nobody knows how to write a `vertex' operator for
string excitations above a non-trivial vacuum, such as the
one existing in QCD. A possible way out is provided
by our second ingredient, namely
conformal invariance. We propose
to use a sigma model technique and request the vanishing
of the corresponding beta functional for the couplings $U(x)$. From
these we shall eventually derive the appropriate
long distance effective action of QCD.

Conformal invariance amounts to demanding that the theory
is independent of the specific conformal factor chosen to describe
the two-dimensional world sheet. While this is
a desirable property of fundamental strings, it need not be
necessarily so (if we look at the QCD string with a magnifying glass
we shall eventually see quarks and gluons, not the string itself!),
so we should rather demand `conformal covariance'. Let us
assume for the time being that conformal invariance is approximately true,
however, and we shall later briefly mention how to move away from this
`zeroth order' approximation.

\section{The model}

In order to obtain the long distance QCD effective action from the
QCD string we follow the strategy used\cite{callan}
to derive Einstein equations from string theory, namely, the non-linear
$\sigma$ model approach.

We couple, in a chiral invariant manner,
the matrix in flavour space $U(x)$, containing  the meson fields, to the
string degrees of freedom while preserving
general covariance in the two dimensional coordinates and conformal
invariance under local scale transformations
of the two-dimensional metric tensor.

The equations of motion for the $U$ field will be obtained from the
condition that the quantum theory must be
conformally invariant, i.e. the
$\beta$ functional for the $U(x)$ couplings must vanish.

Since the string variable $x$ does not contain any flavor dependence,
we have to invent a way to couple it to the background $U$ variable.
We introduce two fermion families living on the
boundary of the string sheet. They carry flavor indices. The action for
the fermions is
\beqn
\frac{1}{2}\int_{\partial\Sigma} d\tau(\bar\psi_L U
\frac{\partial\psi_R}{\partial\tau}- \frac{\partial
\bar\psi_L}{\partial\tau} U \psi_R) + h.c.,
\label{act1}\eeqn
where $\tau$ is the coordinate along the (open) string boundary.
Physically
the labels $R$ and $L$ remind us that these one dimensional Grassmann
variables represent massless quarks of a given chirality moving
along
the ends of the string, and coupled to the external source $U(x)$. Under
$SU(3)_L\times SU(3)_R$ they transform as left- and right-handed fields
do. The above lagrangian  is not unique, but it appears to be
the simplest
one with the desired properties.

The above coupling may appear suprising at first and somewhat ad-hoc.
To see that this is not so, let us expand the non-linear field
$U(x)$, i.e. $U(x)\simeq 1 + i\pi(x)/v + ...$ and retain the first two
terms. The first term just gives rise to a $\theta$-function propagator
which eventually leads to the familiar ordering in the usual
string amplitudes $t_1< t_2< ....$. The second term just provides
(after integrating the fermions out) the usual (tachyonic!) vertex.
In short, if we ignore the non-linearities in the theory we
are back to the usual difficulties.

In order to simplify the calculations, we treat
the couplings $U$ and $U^+$ as independent. The constraint
\beqn
UU^+=1
\eeqn
will be imposed after finding the equations of motion
for an arbitrary
matrix $U$. The reason is simply that we do
not know of an easy way to find the beta function for
constrained coupling
constants.

It is easy to see that the previous action is invariant under general
coordinate
transformations of the two dimensional world sheet by writing it as follows
\beqn
\frac{1}{2}\int_{\partial \Sigma} d\tau \frac{dx^{\mu}}{d\tau}(\bar\psi_L U
\frac{\partial\psi_R}{\partial x^{\mu}}+ \ldots = \frac{1}{2}\int_{\partial
\Sigma} dx^\mu (\bar\psi_L U \frac{\partial\psi_R}{\partial x^{\mu}}
+\ldots  ,\eeqn
where
$\frac{dx^\mu}{d\tau}$ is the tangent vector to the boundary of the
two dimensional surface of the string and the fermions
are treated as scalars under general coordinate transformations.
The fermion action is automatically conformally invariant, because it does
not contain the two dimensional world sheet metric tensor since it can be
written as a line integral.

Notice that $U(x)$ has support only on the boundary of the
string.
The above boundary action has to be supplemented with the
usual bulk action for the string in the conformal gauge. Namely
\beqn
\frac{1}{\alpha^\prime}\int d\sigma d\tau \partial_a x^{\mu} \partial_a
x_{\mu}. \label{act2}\eeqn
Unless otherwise indicated we take $\alpha^\prime=1$.

\section{One loop}

Now we expand $U(x(\tau))$ around a constant background $x_0$ and look
for the potentially divergent
One Particle Irreducible diagrams (OPI).
We classify them according to the number of loops.

The appropriate Feynman rules for the bosonic and fermionic propagators are
\beqn
\langle x^a(\tau) x^b(\tau^\prime)\rangle=\delta^{ab} \Delta_F(\tau
-\tau^\prime)
\eeqn
\beqn
\langle \psi_L(\tau) \bar\psi_R(\tau^\prime)\rangle= U(x_0)^{-1}
\theta(\tau-\tau^\prime)\equiv D_0(\tau-\tau^\prime) \eeqn
Here
$\Delta_F$ is the Feynman propagator for the string coordinate
$x$. The vertices obtained after the expansion of $U(x)$ around $x_0$ lead
to the following Feynman rule for a vertex with $n$ external
$x$ fields
\beqn
V_n=-\frac{1}{n!} \partial_{\mu_1,\mu_2,...\mu_n}U(x_0) \partial_\tau
\eeqn

To renormalize the propagator we have, at this order,
the diagrams shown in figure 1.
These are the only one loop graphs with two fermion legs and zero string
legs.
The calculation is straightforward and we immediately get
for the divergent part of the propagator
\beqn
-\frac{1}{2\epsilon}U^{-1}(\frac{1}{2}\Box{U}
-\partial_{\mu} U U^{-1}\partial^{\mu}U)U^{-1} \theta(\tau-\tau^\prime)
\eeqn
where dimensional regularization
has been used. The $\epsilon$ pole
comes from the singular part of  $\Delta_F(0)$ which also contains the
factor $e^{\phi\epsilon}$. Thus  conformal invariance will be broken at
the one loop level unless
\beqn
\frac{1}{2}\Box{U}-\partial_{\mu} U U^{-1}\partial_{\mu} U=0
\eeqn
These are the equations of motion of the $U$ field. Later we will
supplement them with the unitarity constraint.
The fermion propagator can be made finite
by using minimal subtraction and redefining accordingly
\beqn
U(x_0)^{-1}\to U(x_0)^{-1}+\delta^{(2)} U(x_0)^{-1}.
\eeqn
with $\delta^{(2)} U^{-1}$ given by
\beqn
\delta^{(2)} U^{-1}=
\frac{1}{2\epsilon}U^{-1}(\frac{1}{2}\Box{U}
-\partial_{\mu} U U^{-1}\partial^{\mu}U)U^{-1}.
\eeqn

Next we turn to the vertices with one $x$- and
two $x$-fields. The relevant diagrams are shown in
figures 2 and 3, respectively. A direct calculation,
and the use of
\beqn
\delta U = - U \delta U^{-1} U,
\eeqn
shows
that the counterterm needed to cancel the divergent part for the former is
just \beqn
\delta^{(2)} V_1=-\partial_{\mu} \delta U \partial_\tau,
\eeqn
while the conterterm of the latter is
\beqn
\delta^{(2)} V_2=-\frac{1}{2} \partial_{\mu}\partial_{\nu} \delta U \partial
_\tau. \eeqn
These expressions will be needed for the two loop calculation.

The vanishing of the beta functional for $U(x)$ can be obtained as
the Euler-Lagrange variation of a given action $\tilde S$. The
true action will however be
\beqn
S=\tilde S+\int d^nx\ {\rm tr}(\lambda(x)(U(x)U^+(x)-1))
\eeqn
It is easy to see that the variation of $S$ produces the
equations of motion
\beqn
U\Box{U^+}-\Box{U}U^+=0,
\eeqn
which are the ones derived from the chiral lagrangian at lowest order (see
appendix).

Thus we have succeded in deriving a long-distance effective action
for QCD with all the required properties, at least at this order.

\section{Two loops}

 In order to compute the two-loop corrections to the fermion propagator in
eq.13 one has to consider first the diagrams in figure 4. The total result
amounts to
\beqn
-\frac{1}{2}D_0(\tau-\tau^\prime)\Delta^2_F(0)T
\eeqn
where $T$ is given by
\beqn
T=\frac{1}{4}O_1-O_2-O_3-O^\prime_3+2O_4+O_5-2O_6-2O_7+2O_8+2O^\prime_8
\eeqn
and the $O_i$ operators are defined as
\beqn
O_1 & = & \Box ^2UU^{-1}    \nonumber    \\
O_2 & = &
\partial_{\mu}\partial_{\nu}UU^{-1}\partial^{\mu}\partial^{\nu}UU^{-1}
\nonumber   \\
O_3 & = & \partial_{\mu}UU^{-1}\Box\partial^{\mu}UU^{-1}    \nonumber \\
O^\prime_3 & = & \Box \partial_{\mu}UU^{-1}\partial^{\mu}UU^{-1} \nonumber
\\
O_4 & = &
\partial_{\mu}UU^{-1}\partial^{\mu}\partial^{\nu}UU^{-1}\partial_{\nu}UU^{-1} \nonumber  \\
O_5 & = & \partial_{\mu}UU^{-1}\Box UU^{-1}\partial^{\mu}UU^{-1}
\nonumber \\
O_6 & = & \partial_{\mu}UU^{-1}\partial_{\nu}UU^{-1}\partial^{\mu}UU^{-1}
\partial^{\nu}UU^{-1}   \nonumber  \\
O_7 & = & \partial_{\mu}UU^{-1}\partial_{\nu}UU^{-1}\partial^{\nu}UU^{-1}
\partial^{\mu}UU^{-1}    \nonumber  \\
O_8 & = &
\partial_{\mu}\partial_{\nu}UU^{-1}\partial^{\mu}UU^{-1}\partial^{\nu}
UU^{-1}    \nonumber    \\
O^\prime_8 & = & \partial_{\mu}UU^{-1}\partial_{\nu}UU^{-1}
\partial^{\mu}\partial^{\nu}UU^{-1}
\eeqn
Notice the appeareance of four derivatives in the above expressions. The
two-loop calculation is the relevant one for the ${\cal O}(p^4)$
terms of the chiral lagrangian.

In addition we have also the contribution coming from the counterterm
diagrams $I$, $II$ and $III$ appearing in figure 5
\beqn
D_0(\tau-\tau^\prime)\frac{1}{2\epsilon}\Delta_F(0)(D_{I}+D_{II}+D_{III})
\eeqn
where
\beqn
D_{I} & = & \frac{1}{2}O_5-O_7  \nonumber  \\
D_{II} & = & -\frac{1}{2}(O_3+O^\prime_3)+2O_4-2O_6+O_8+O^\prime_8
\nonumber       \\
D_{III} & = & \frac{1}{4}O_1-\frac{1}{2}(O_3+O^\prime_3)-O_2
+O_8+O^\prime_8-O_7+\frac{1}{2}O_5
\eeqn
Thus the counterterm contribution is also proportional to the $T$ operator.

Finally, the complete two loop divergent part of the fermion propagator
is
\beqn
  \frac{1}{8\epsilon^2}D_0(\tau-\tau^\prime) T
\eeqn
so that no simple $\epsilon$ pole appears. The two-loop
fermion propagator is made finite with the help of the
counterterm
\beqn
\delta^{(4)}U^{-1} = - \frac{1}{8\epsilon^2}U^{-1} T
\eeqn

The absence of simple poles at the two loop level implies in minimal
subtraction that the two loop contribution to the beta functional is
zero. Thus there is no net contribution to the equation of motion
at order $(\frac{1}{\alpha^\prime})^2$. (Notice that $\Delta_F$ actually
contais a $\frac{1}{\alpha^\prime}$ factor.) Therefore the requirement
of conformal invariance
implies $L_1=L_2=L_3=0$.

\section{Discussion}

Is it possible to understand the vanishing of the ${\cal O}(p^4)$
coefficients? Here we provide a tentative argument.

One must first realize that the matrix $U(x)$ is dimensionless and thus
cannot solely depend on $x^\mu$, some dimensional quantity is required.
Let us call this quantity $v$ (of course, $v=F_\pi$, but we do not need
to know this at his point). Then the full action we have written
(eqs. (\ref{act1}) plus (\ref{act2})) is trivially invariant under the
following set of transformations
\beqn
x\to e^t x, \qquad v \to e^{-t} v, \qquad
\alpha^\prime \to e^{2t}\alpha^\prime,
\eeqn
One may say that this is not really an invariance since we
change both fields (which are integrated over) and couplings (which are
not).
In fact if this were the full story this would imply nothing for
$L_1$, $L_2$  and $L_3$ since, on dimensional grounds and counting powers
of
$\alpha^\prime$, they must be of the form  $L_i \propto \alpha^\prime v^2$.

However, it turns out that conformal invariance would imply that the
invariance is stronger, since the change in $\alpha^\prime$ can
be absorbed by the shift
$\phi \to \phi-2t/\epsilon$ once the theory is regulated by continuing
it away from 2 dimensions. Conformal invariance would guarantee
independence of the conformal factor and thus the real invariance of the
theory would be
\beqn
x\to e^t x, \qquad v \to e^{-t} v.
\eeqn
This would imply $L_1=L_2=L_3=0$, in fact all higher order coefficients
appear to vanish for exactly the same reasons if this argument holds.
However, the argument is only tentative, since the theory is after all
not conformally invariant in all sectors since we are away from the
critical dimension. The calculation we have just presented is what
really settles the issue.

In real QCD the ${\cal O}(p^4)$ coefficients are known to be of
order $10^{-3}$.  It appears thus that assuming conformal
invariance of the string propagating in a chirally-non invariant vacuum is
not such a bad approximation.
(Of course the smallness of the $L_i$ can be understood on other grounds,
but these have nothing to do with the string.)

In large $N$ QCD (the theory the string is supposedly reproducing)
$F_\pi$ is known to be of order $\sqrt{N}$. Hence quantum loop
effects are absent in the chiral lagrangian. On the other hand,
all resonances are narrow. Consequently
$L_i\sim\sum_n f^2_n/m_n^2$. If no higher spins are included, then
in the large $N$ limit $L_i=0$. This is precisely what we get. The
consistency with large $N$ QCD is striking.

To obtain more physical values for the $L_i$ we should extend
our program to include higher spin external fields (such as vectors
and axial vectors), in a way similar to what was done
in \cite{labas} for the open bosonic string (perturbed around the
usual vacuum). Demanding conformal invariance of the effective action
via the vanishing of the beta functionals would lead to a system of
coupled differential equations for all these degrees of freedom, from
where an effective lagrangian
containing pions, vector mesons, etc could be inferred.
The subsequent integration of the higher spin states would
give non-zero values for the $L_i$.

Of course another way to obtain
non-zero values for $L_1$, $L_2$ and $L_3$ is to give up
conformal invariance. For that one must
use a string action which manifestly breaks
conformal invariance. The simplest possibility
is to include the extrinsic curvature term. The $L_i$
would then be  get a term proportional to the rigidity coefficient.
Perhaps this would be the appropriate way to include
the $1/N$ corrections. A hint in this direction come from
the well-known fact that integration of fermions (suppressed by $1/N$)
in the supersymmetric string leads, amongst other things, to
the appeareance of extrinsic curvature.

In conclusion, we have seen that while attempting to build
string operators describing excitations above the `right'
physical vacuum is probably hopeless, the sigma model
approach bypasses this difficulty by determining which are
the `classical' backgrounds where propagation of the
bosonic string is consistent. The `perturbative' vacuum
built of tachyons, massless vectors, etc. is a consistent
one (from the string point of view, not of QCD, of course!).
But a chirally non-invariant vacuum with massless scalars
(interacting with a non-linear lagrangian) is consistent too and certainly
a lot more physical. The tachyon is gone. We are perfectly well aware that
the string action
we have used is a sick one and cannot be used beyond the string tree level,
but, as said, this is not really a fundamental difficulty for
an effective theory. We have found that conformally symmetric
string actions are a good starting point, contrary to a common
belief. Perhaps the old ideas of Cremmer and Scherk could
be finally implemented following the present lines.

\bigskip\bigskip

\section*{Acknowledgements}
\bigskip

We thank
A.Andrianov for multiple discussions. This work
was initiated during the visit of one of the authors
to the Departamento de F\'\i sica of the
Universidad Cat\'olica de Chile, whose hospitality
is gratefully acknowledged.
We acknowledge the financial support from
grants CICYT AEN98-0431 and AEN96-1634, CIRIT 1996SGR00066,
and, specially, from the `Programa de Cooperaci\'on
con Iberoam\'erica'. J.A. is partially supported by the project Fondecyt
1980816.

\vfill
\eject

\section*{Appendix}

At long distances QCD is described by the chiral lagrangian. This
is an effective lagrangian organized in powers of momenta
(see e.g. \cite{dobado} for a general discussion)
\beqn
{\cal L}_{\rm eff}= {\cal L}^{(2)} + {\cal L}^{(4)}+\ldots
\eeqn
where
\beqn
{\cal L}^{(2)}=\frac{F_\pi^2}{4}{\rm tr} \partial_\mu U \partial^\mu
U^\dagger ,
\eeqn
\beqn
{\cal L}^{(4)}= L_1 ({\rm tr} (\partial_\mu U \partial^\mu U^\dagger))^2
+
L_2 {\rm tr} (\partial_\mu U \partial_\nu U^\dagger)
{\rm tr} (\partial^\mu U \partial^\nu U^\dagger)
+
L_3 {\rm tr} (\partial^\mu U \partial^\nu U^\dagger \partial_\mu U
\partial_\nu U^\dagger).
\eeqn
The matrix $U=\exp {\rm i}\tau^a \pi^a/ 2F_\pi$ collects the Goldstone
bosons associated to the $SU(3)_L\times SU(3)_R\to SU(3)_V$ breaking.

The experimental values for these low-energy constants are (at the $m_\eta$
scale): $L_1=(0.65\pm 0.28)\times 10^{-3}$,
$L_2=(1.90\pm 0.26)\times 10^{-3}$ and $L_3=(-3.06\pm 0.92)\times 10^{-3}$.
They are, generally speaking, well accounted for by either the chiral
quark model or vector meson saturation (the latter, incidentally, explains
the good agreement with the Lovelace-Shapiro amplitude predictions).

\vfill\eject

\vfill\eject

\begin{figure}
\vspace{1.0in}
\includegraphics{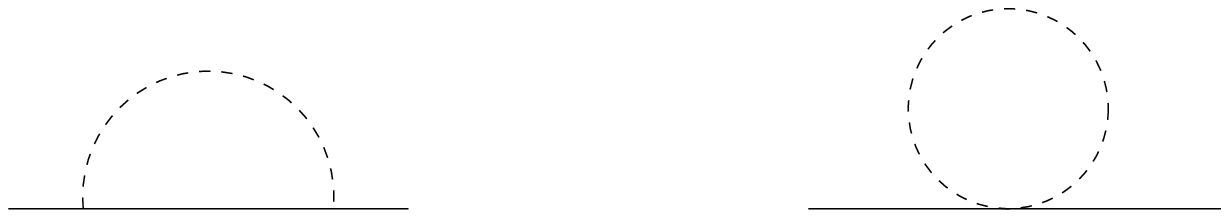}
\caption{One-loop diagrams for the propagator. }
\end{figure}

\begin{figure}
\vspace{2.5in}
\includegraphics{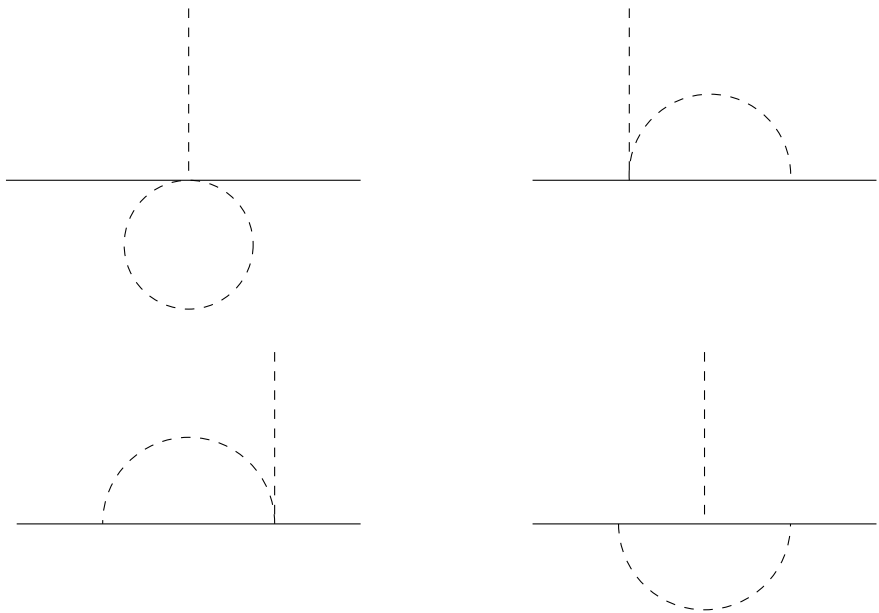}
\caption{One-loop diagrams for the vertex with one $x$-field. }
\end{figure}

\begin{figure}
\vspace{4in}
\includegraphics{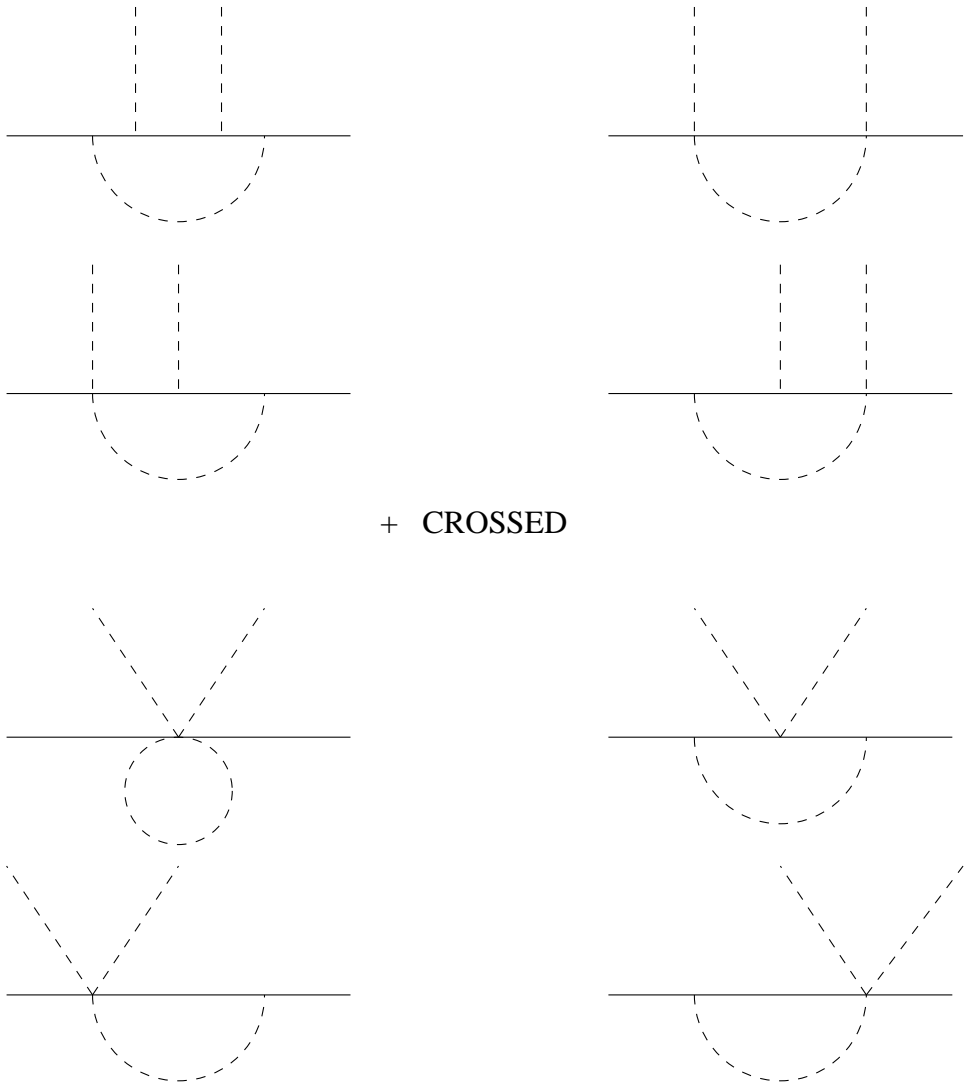}
\caption{One-loop diagrams for the vertex with two $x$-fields. }
\end{figure}

\begin{figure}
\vspace{4in}
\includegraphics{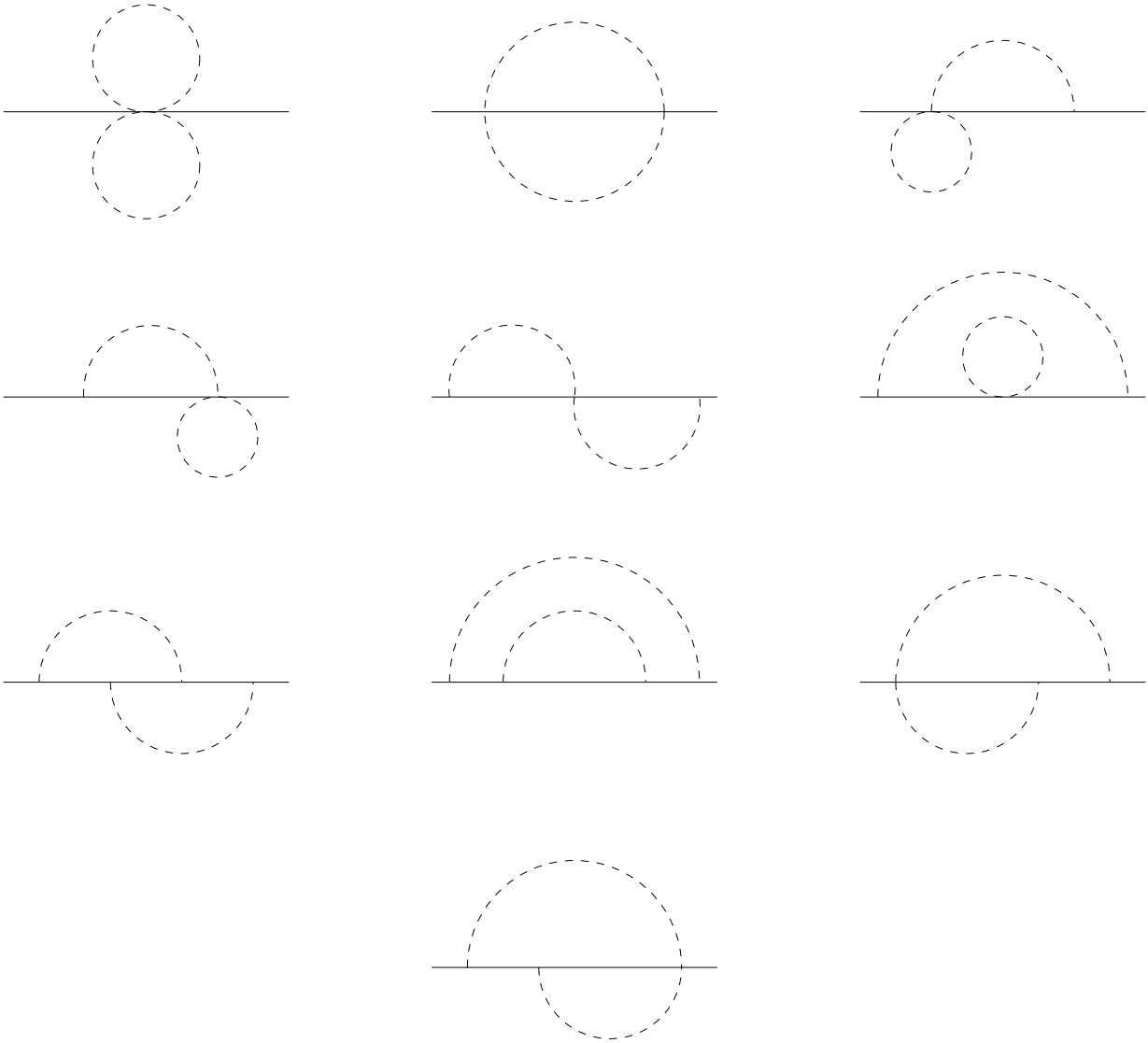}
\caption{Two-loop diagrams for the propagator. }
\end{figure}

\begin{figure}
\vspace{4.0in}
\includegraphics{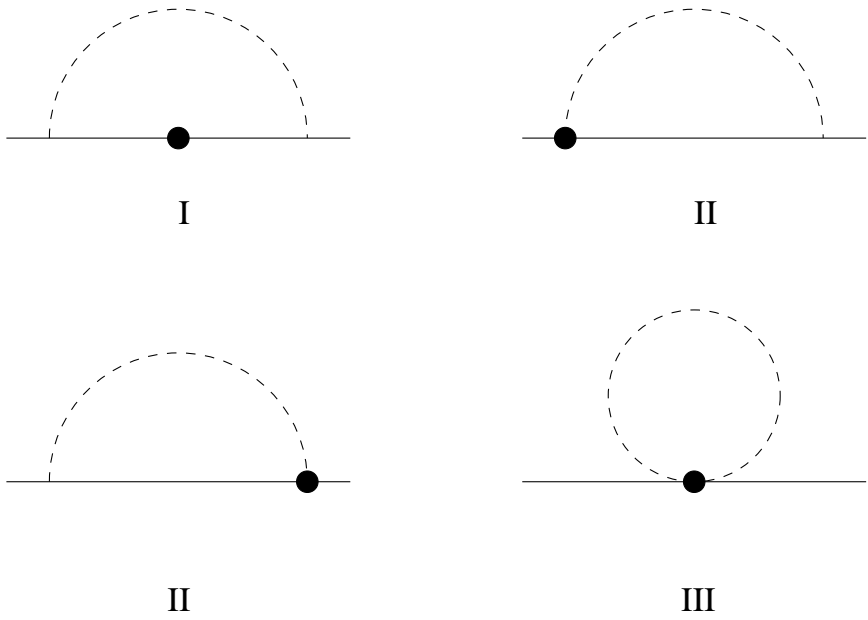}
\caption{One-loop counterterms. }
\end{figure}

\end{document}